\documentclass[prl,showpacs,preprintnumbers,twocolumn,amsmath,amssymb,superscriptaddress]{revtex4-1}

\usepackage{graphicx}
\usepackage[latin1]{inputenc}
\usepackage{amsfonts}
\usepackage{amsmath,amssymb}
\usepackage{bm}
\usepackage{ulem}
\usepackage{color}

\begin{document}
\graphicspath{{images/}}

\title{Fluctuation-induced current from freestanding graphene: toward nanoscale energy harvesting}
\author{P. M. Thibado$^{*}$}
\affiliation{Department of Physics, University of Arkansas, Fayetteville, Arkansas, 72701, USA.}
\author{P. Kumar}
\affiliation{Department of Physics, University of Arkansas, Fayetteville, Arkansas, 72701, USA.}
\author{Surendra Singh}
\affiliation{Department of Physics, University of Arkansas, Fayetteville, Arkansas, 72701, USA.}
\author{M. Ruiz-Garcia}
\affiliation{Department of Physics, University of Pennsylvania, Philadelphia, Pennsylvania, 19104, USA}
\author{A. Lasanta}
\affiliation{G. Mill\'an Institute for Fluid Dynamics, Nanoscience and Industrial Mathematics and Department of Mathematics, Universidad Carlos III de Madrid, 28911 Legan\'es, Spain}
\affiliation{Departamento de \'Algebra. Facultad de Educaci\'on, Econom\'\i a y Tecnolog\'\i a de Ceuta, Universidad de Granada, Cortadura del Valle, s/n. E-51001 Ceuta, Spain}
\author{L. L. Bonilla$^{**}$,}
\affiliation{G. Mill\'an Institute for Fluid Dynamics, Nanoscience and Industrial Mathematics and Department of Mathematics, Universidad Carlos III de Madrid, 28911 Legan\'es, Spain}
\affiliation{Courant Institute for Mathematical Sciences, New York University, 251 Mercer St., New York, NY 10012, USA\\
$^*$Corresponding author. E-mail: thibado@uark.edu\\
 $^{**}$Corresponding author. E-mail: bonilla@ing.uc3m.es}
\date{\today}

\begin{abstract}
At room temperature, micron-sized sheets of freestanding graphene are in constant motion even in the presence of an applied bias voltage. We quantify the out-of-plane movement by collecting the displacement current using a nearby small-area metal electrode and present a Langevin model for the motion coupled to a circuit containing diodes. The system reaches thermal equilibrium and the rates of heat, work, and entropy production tend quickly to zero. However, there is power generated by graphene which is equal to the power dissipated by the load resistor. The exact power formula is similar to Nyquist's noise power formula, except that the rate of change of diode resistance significantly boosts the output power, and the movement of the graphene shifts the power spectrum to lower frequencies.
\end{abstract}

\maketitle

Freestanding, two-dimensional (2D) crystalline membranes exhibit unique out-of-plane motion \cite{cas09}. When relaxed, sheets of freestanding graphene feature a rippled morphology, in which adjacent regions alternate between concave and convex curvature \cite{mey07}. The origin of these nanometer-sized ripples is still an open question \cite{fas07,abe07}. Theoretical work points to electron-phonon coupling as the source because it suppresses long-wavelength bending rigidity and enhances off-plane fluctuations \cite{gaz09,san11,gui14}. In a state of thermal equilibrium, Guinea {\em et al} derived a system of equations for the height of a graphene membrane including auxiliary stress and curvature fields \cite{gui14}. Within this perturbative formulation of quantum statistical mechanics, circular graphene membranes spontaneously buckle below a critical temperature and above a critical radius \cite{bon16}. Numerical studies of static rippling in a membrane coupled with Dirac fermions show a phase transition from flat to rippled morphology \cite{cea19prl,cea19prb}.

To date, no studies have been undertaken of dynamic fluctuations using a Hamiltonian that includes Dirac electrons, elasticity, and the electron-phonon interaction. Early phenomenological studies modeled the electron-phonon interaction by coupling point particles at the nodes of a hexagonal lattice to Ising spins that undergo Glauber dynamics \cite{bon12,bon12jstat}. The spins exchange energy with a thermal bath, their dynamics show rippling, and their interaction with the membrane drives the whole system to equilibrium \cite{mrg15}. 

Experiments by Ackerman {\em et al} have measured the out-of-plane motion of atoms in freestanding graphene using scanning tunneling microscopy (STM) \cite{ack16}. They show that single atoms in the membrane experience Brownian motion with sporadic large jumps that are typical of L\'evy processes \cite{ack16,van03}. Rare jumps in the height of the graphene atoms correspond to coherent inversions of the curvature of the ripples upon which the atoms sit. This is consistent with both molecular dynamics \cite{ack16} and spin-membrane Glauber dynamics \cite{mrg15,sch15,mrg16}. 

In the present study, graphene was commercially grown on Ni and transferred to a 2000-mesh, ultrafine copper grid featuring a lattice of square holes (each 7.5 $\mu$m wide) and bar supports (each 5 $\mu$m wide). Excess graphene bonds to the side wall \cite{bun08}. Scanning electron microscope images confirmed 90\% coverage of the grid \cite{xu12}. 

An Omicron ultrahigh vacuum STM (base pressure $10^{-10}$ mbar) operated at room temperature was used. Graphene film was mounted toward the sample plate on standoffs, allowing the STM tip to approach through the grid holes. The entire STM chamber rests on an active, noise-cancelling, vibration isolation system. It is powered using a battery bank with an isolated ground to achieve exceptionally low mechanical and electrical noise.

\begin{figure}[h]
\begin{center}
\includegraphics[width=8cm]{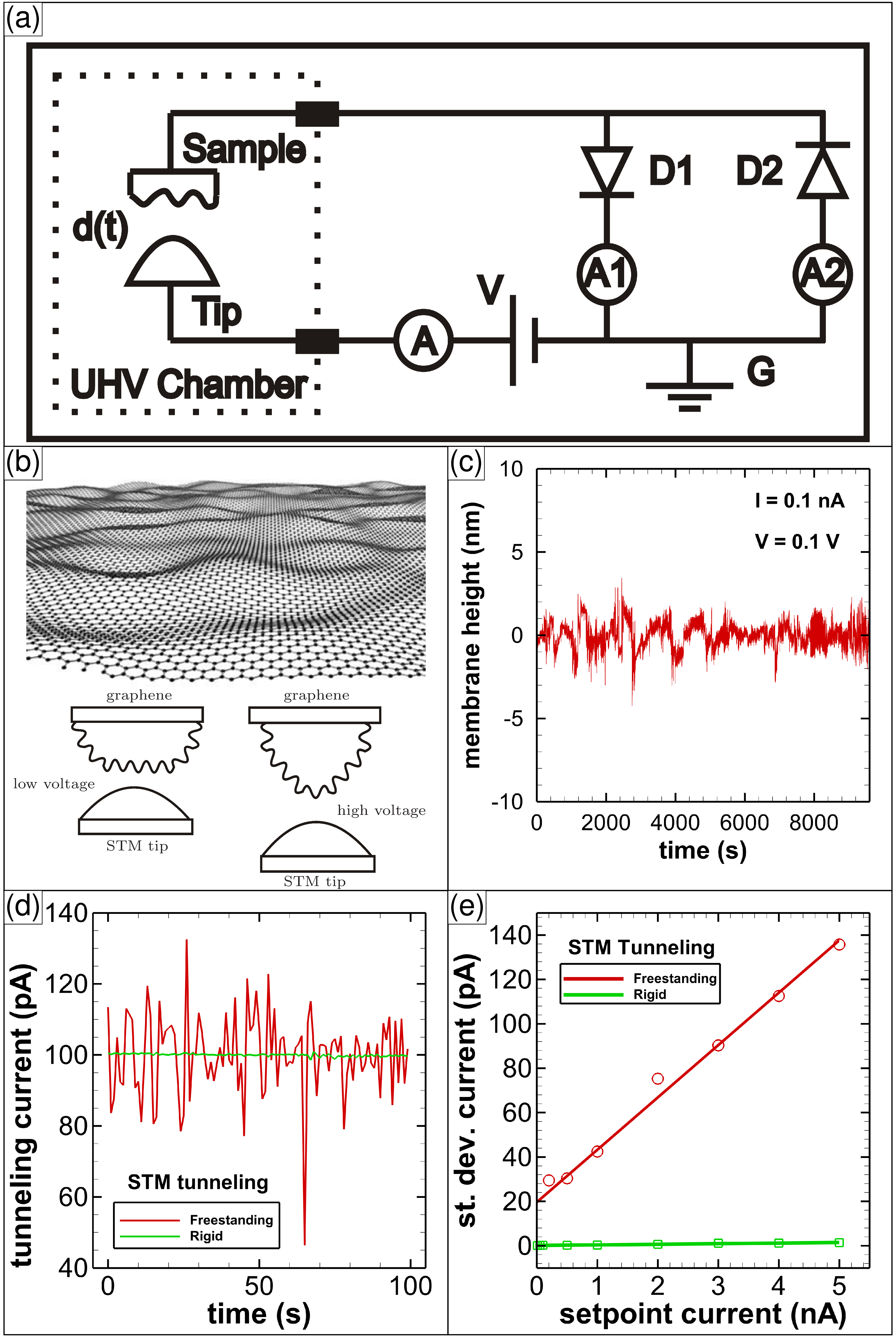}
\end{center}
\caption{STM data sets acquired when STM tip is tunneling electrons. (a) Circuit diagram showing STM tip, sample, bias voltage, ammeters, and diode arrangement. (b) Sketch of graphene sheet in rippled state and illustrations of graphene shape changes. (c) Height fluctuations of graphene. (d) STM tunneling current vs time for freestanding and rigid graphene. (e) Standard deviation of tunneling current vs setpoint current for freestanding  and rigid graphene. \label{fig1}}
\end{figure}

The STM tip-sample junction is incorporated into the circuit shown in Fig.~\ref{fig1}(a). The sample is isolated from ground and connected to two diodes \cite{xu06}. The tip-sample junction acts as a variable capacitor \cite{abd16,phi77,gre05,odo09}. The tunneling current, diode 1 current (D1C), and diode 2 current (D2C) are monitored simultaneously. This diode arrangement is used for energy harvesting, but here we use it to isolate the graphene-induced current from the battery current. At a tip-sample distance of 2 nm or less, tunneling electrons dominate the current; for larger distances, displacement current dominates. 

An illustration of rippled graphene and voltage-induced shape changes is shown in Fig.~\ref{fig1}(b). When the bias voltage increases the graphene is stretched and the STM tip moves with the graphene. We attribute this movement to the electric force, which was characterized in our earlier study \cite{xu13}. A typical constant-current point-mode STM measurement of the membrane height in time is shown in Fig.~\ref{fig1}(c). During this experiment the STM tip only moves vertically. Note the enormous size of the movement as compared to atomic corrugations of less than 0.1 nm. The tunneling current in time is shown in Fig.~\ref{fig1}(d) for both rigid graphene (i.e., graphene on copper) and freestanding graphene. For the freestanding sample, the average current is the same as the rigid sample, but the fluctuations are 100 times larger (10 pA vs. 0.1 pA). The result shown in Fig.~\ref{fig1}(d) is independent of the applied bias voltage (up to 3 V) and feedback gain setting. As the setpoint current (SPC) increases, the standard deviation also increases, as shown in Fig.~\ref{fig1}(e). We attribute this to sample heating, which was also previously characterized \cite{xu13}. When extrapolated to zero tunneling current, the fluctuations still contribute 20 pA of displacement current.

\begin{figure}[ht]
\begin{center}
\includegraphics[width=8cm]{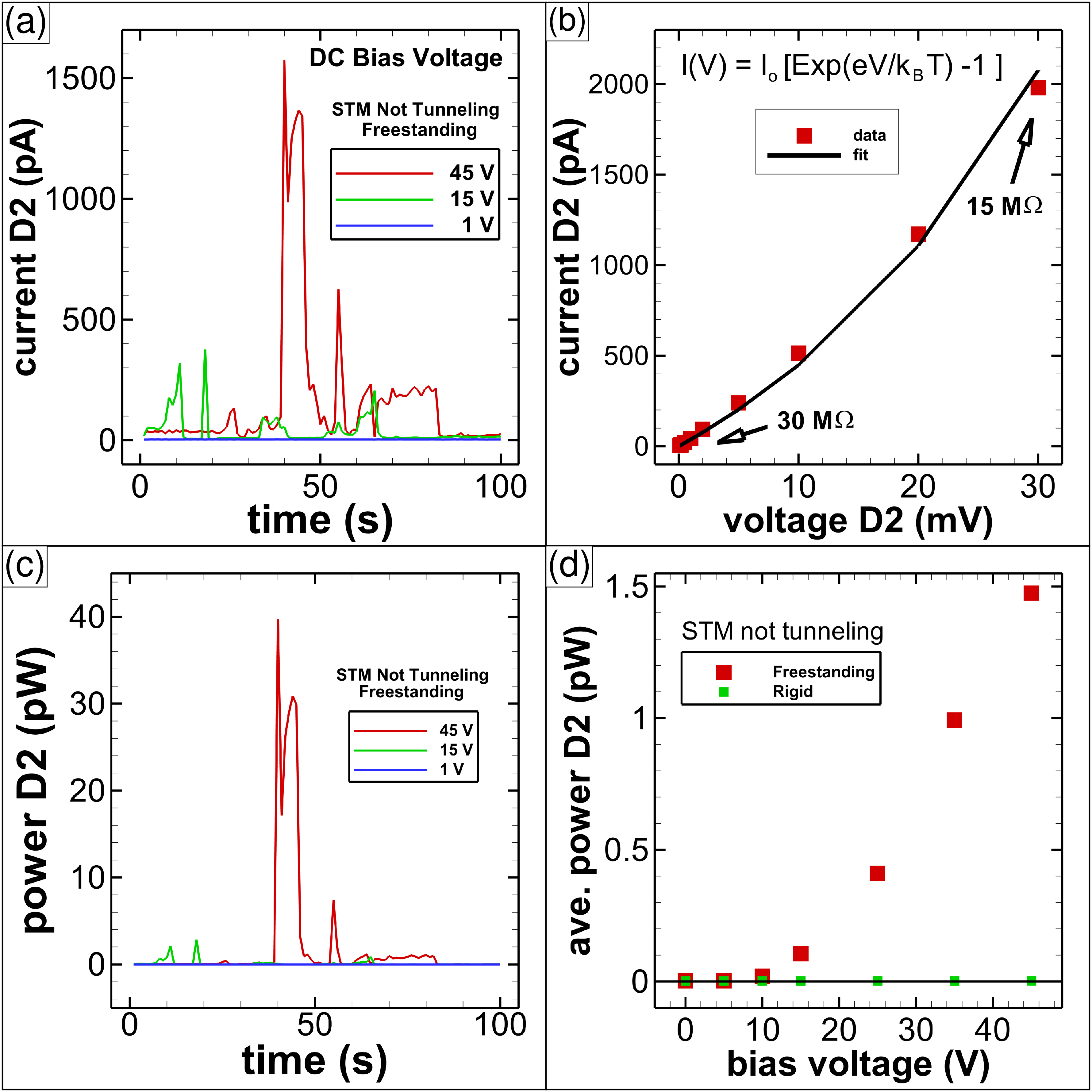}
\end{center}
\caption{STM data sets acquired when the STM tip is not tunneling electrons. (a) Current through diode 2 versus time for voltages $V=1,\, 15,\, 45$ V. (b) Average current vs voltage through diode 2. (c) Power through diode 2 vs time for different voltages. (d) Average  power through diode 2 vs voltage bias $V$.  \label{fig2}}
\end{figure}

To measure the displacement current at zero tunneling current, we incrementally backed the STM tip away from the sample using the coarse motion stage until the distance was too great for electrons to tunnel through the vacuum barrier. In this position, the SPC is at 50 nA, thereby using the feedback circuit to keep the STM tip stationary (i.e., fully forward). Once in position, we applied a DC bias voltage and recorded the D2C in time, as shown in Fig.~\ref{fig2}(a). At one volt, no current is induced, but at 15 V and 45 V, we systematically observed a spiky, time-dependent D2C. If and when a tunneling current was detected, we repositioned the tip further from the graphene. This is done by taking coarse motion steps away from the graphene or by fine adjustment of the lateral position of the tube scanner.

The low current I-V characteristics of the diode are shown in Fig.~\ref{fig2}(b), with resistances labeled. The power dissipated in diode 2 was then calculated, as shown in Fig.~\ref{fig2}(c), which reaches 40 pW.   The average power for a large number of freestanding graphene and rigid graphene data sets acquired across this sample and other identically prepared samples is shown in Fig.~\ref{fig2}(d). The lack of current for the rigid sample confirms that contamination and electron field emissions are not the sources of the D2C.

\begin{figure}[h]
\begin{center}
\includegraphics[width=8cm]{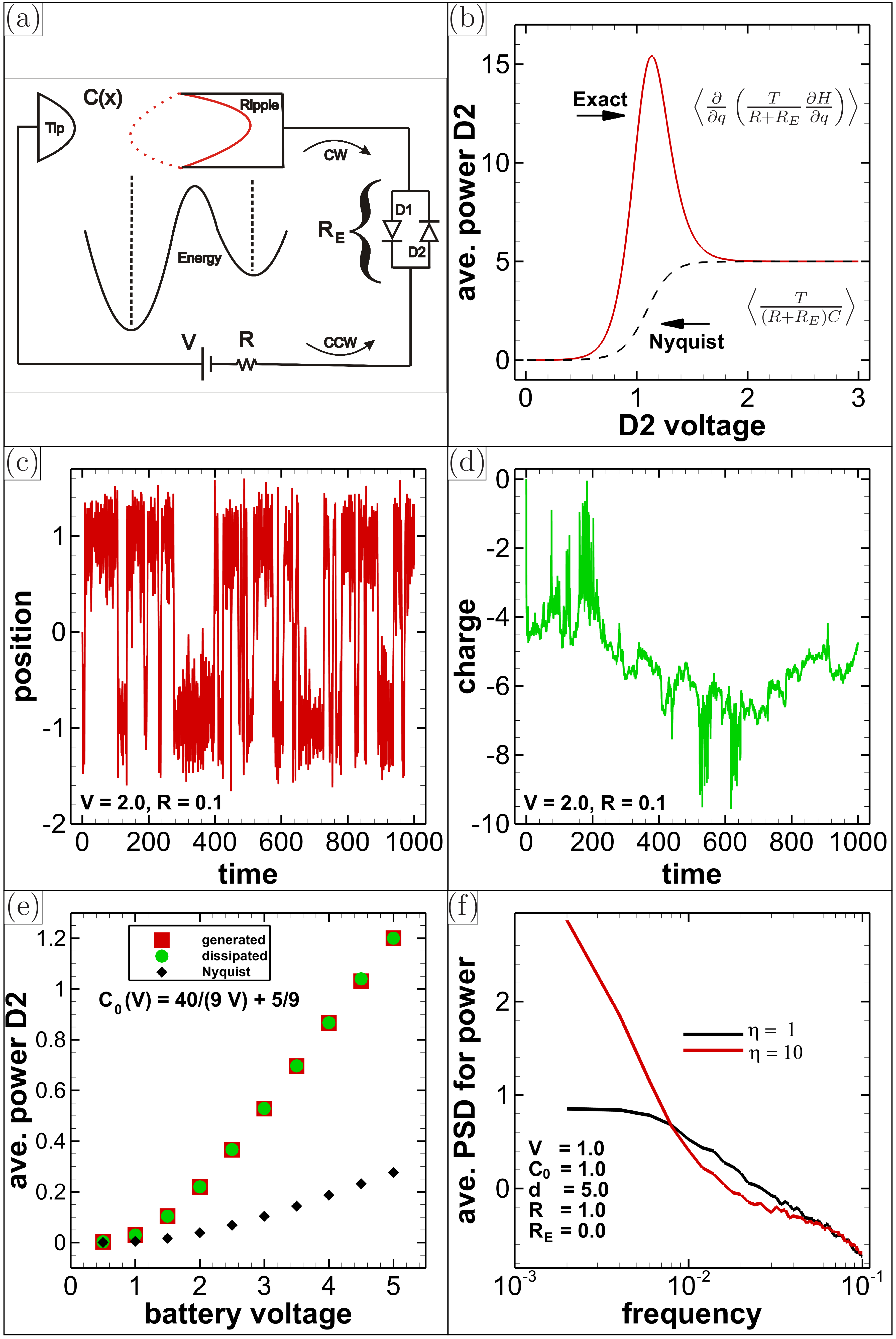}
\end{center}
\caption{Langevin equation simulation results for a circuit with diodes and resistor. (a) Sketch of circuit with energy barrier diagram. (b) Noise power vs diode voltage showing the power enhancement over Nyquist's formula. (c) Height of graphene ripple $x$ and (d) charge $q$ vs time. (d) Average counterclockwise power versus battery voltage. (f) Average power spectrum density of power vs frequency. }
\label{fig3}
\end{figure}

These data suggest that electrical work is done on D2 by the motion of the graphene even though it is held at a single temperature (i.e., room temperature). Work can be done while in thermodynamic equilibrium, and a deeper understanding of this will shed light on potential methods for non-equilibrium energy harvesting \cite{mar16}. To achieve this goal, we developed the model illustrated in Fig.~\ref{fig3}(a). The carbon atom closest to the STM tip sits over a ripple, which fluctuates between convex and concave curvature. We model this situation as a Brownian particle in a double-well potential in contact with a thermal bath at temperature $T$ (in units of energy). The damping force and thermal noise satisfy the Einstein relation. The STM tip and sample act as a capacitor of variable capacitance $C(x)=C_0/(1+x/d)$, where $d+x(t)$ and $x(t)$ ($x\ll d$) are the instantaneous distance between STM tip and sample, and the vertical position of the carbon atom measured with respect to the flat configuration of the graphene membrane, respectively. If the instantaneous charge and voltage drop of the tip sample capacitor are $q(t)$ and $u(t)$, respectively, then the electrostatic force exerted over the particle is $qu/[2(d+x)]=q^2/(2C_0d)$. The equation for the charge $q(t)$ follows from Kirchhoff's law. Therefore, the coupled systems of particle and circuit satisfy the Langevin-Ito equations
\begin{eqnarray}
&&\dot{x}=v,\nonumber\\
&&\dot{v}=-\eta v- U'(x) -\frac{q^2-C_0^2V^2}{2C_0d} + \sqrt{2T\eta}\, \xi_v(t), \nonumber\\
&&\dot{q}=\frac{\partial}{\partial q}\!\left(\frac{T}{\mathcal{R}}\right)\!-\frac{V+u}{\mathcal{R}}+\sqrt{\frac{2T}{\mathcal{R}}}\, \xi_q(t), \quad u= \frac{q}{C(x)},\quad\label{eq1}
\end{eqnarray}
where $U(x) = x^4-2x^2$ is a double well potential, $\frac{C_0V^2}{2d}$ is a constant tension due to graphene stretching, $\mathcal{R}=R+R_E$ is the total resistance, $\frac{1}{R_E} = \frac{2 I_0}{u_D} \sinh \frac{u_D}{T_e}$ is the equivalent resistance of the diodes, $u_D$ is the voltage drop across the diodes, $T_e = \frac{T}{e}$, $\frac{\partial}{\partial q}(\frac{T}{\mathcal{R}})$ is the noise-induced drift correction, and $\xi_v$ and $\xi_q$ are zero-mean independent and identically distributed white noise with delta correlations $\langle \xi_i(t)\xi_j(t')\rangle=\delta_{ij}\delta(t-t')$ $i,j=v,q$. The circuit equation has Nyquist noise at temperature $T$, which is set to the same temperature as the graphene ripple. The $\frac{\partial}{\partial q}(\frac{T}{\mathcal{R}})$ term guarantees detailed balance and that the overall system reaches thermal equilibrium at temperature $T$. To see this, we start from a master equation for the electron probability density with transition rates obeying detailed balance \cite{lan62,vk60,sok98,sok99}. We choose the transition probability as $T(i_{D1}+i_{D2})/(e^2u_D)=T/(e^2\mathcal{R})$, which is  consistent with Kirchhoff's law for the currents. In the continuum limit, the master equation then becomes the Fokker-Planck equation associated with Eq.~\eqref{eq1}. 

The Hamiltonian for the system shown  in Fig.~\ref{fig3}(a) is 
\begin{eqnarray}
&& \mathcal{H}(x,v,q)=\frac{v^2}{2}+U(x)+\frac{q^2}{2 C(x)} + qV - \frac{C_0V^2}{2} \frac{x}{d}, \label{eq2}
\end{eqnarray}
and the equilibrium probability density is $e^{-\mathcal{H}/T}/Z$ where $Z$ is a normalization constant.

From the point of view of the graphene ripple, represented by the particle in Eq.~\eqref{eq1}, the circuit is an external system that exerts work on it. The heat produced by friction and noise forces is then \cite{sekimoto}
\begin{eqnarray}
d'Q&=&\left(-\eta v+ \sqrt{2T\eta}\,\xi_v(t)\right)\!\circ dx(t)\nonumber\\
&=&d\mathcal{H}(x,v,q)-\frac{\partial\mathcal{H}}{\partial q}\circ dq(t),\label{eq3}
\end{eqnarray}
in which $q=q(t)$ is an external parameter, $d'Q>0$ if heat is absorbed by the particle, and the product $\circ$ with differentials is taken in the Stratonovich sense. The Stratonovich product on the first line of Eq.~\eqref{eq3} can be converted into an Ito product: $\xi_v(t)\circ dx(t)=v(t)dw_v(t)+\frac{1}{2}\sqrt{2T\eta}\, dt$, where $dx=v\, dt$ and $dw_v(t)= \xi_v(t)\, dt$ with $(dw_v)^2=dt$ is the differential of the Wiener process corresponding to the white noise $\xi_v(t)$. Then
\begin{eqnarray}
d'Q=\eta\, (T- v^2)\, dt+ \sqrt{2T\eta}\, v\, dw_v(t).\label{eq4}
\end{eqnarray}
The average of the noise term in Eq.~\eqref{eq4} vanishes due to the nonanticipative character of the Ito product, thereby yielding
\begin{eqnarray}
\left\langle\frac{d'Q}{dt}\right\rangle\!=\eta\, (T- \langle v^2\rangle).  \label{eq5}
\end{eqnarray}
This average heat flux becomes zero in equilibrium because of the equipartition theorem.

From the second line in Eq.~\eqref{eq3}, the first law of thermodynamics is obtained,
\begin{eqnarray}
d\mathcal{H}(x,v,q)= d'Q+d'W,   \label{eq6}
\end{eqnarray}
where the work exerted on the particle by the circuit is
\begin{eqnarray}
&&d'W\!=\!\frac{\partial\mathcal{H}}{\partial q}(x,v,q)\!\circ\! dq\nonumber\\
&&=\frac{\partial\mathcal{H}}{\partial q}\!\left[\frac{\partial}{\partial q}\!\left(\frac{T}{\mathcal{R}}\right)\!-\frac{1}{\mathcal{R}} \frac{\partial\mathcal{H}}{\partial q}\right] dt + \sqrt{\frac{2\, T}{\mathcal{R}}}\frac{\partial\mathcal{H}}{\partial q}\!\circ\! dw_q\nonumber\\
&&=\left[\frac{\partial}{\partial q}\!\left(\frac{T}{\mathcal{R}}\frac{\partial\mathcal{H}}{\partial q}\right)\!-\frac{1}{\mathcal{R}}\!\left(\frac{\partial\mathcal{H}}{\partial q}\right)^2\right] dt+ \sqrt{\frac{2\, T}{\mathcal{R}}}\,\frac{\partial\mathcal{H}}{\partial q}\, dw_q.\label{eq7}
\end{eqnarray}
The Stratonovich product is converted to an Ito product on the last line of this expression. From Eq.~\eqref{eq2}, the average power absorbed by the particle is
\begin{eqnarray}
\!\left\langle\frac{d'W}{dt}\!\right\rangle\!=\!\left\langle\frac{\partial}{\partial q}\left(\frac{T}{\mathcal{R}}\frac{\partial\mathcal{H}}{\partial q}\right)\!\right\rangle\!-\!\left\langle\frac{1}{\mathcal{R}}\left(\frac{\partial\mathcal{H}}{\partial q}\right)^2\right\rangle\!.  \quad\label{eq8}
\end{eqnarray}
Using the equilibrium probability density to calculate the average and integrating by parts, the average power absorbed by the particle is found to be zero. The voltage drop $\frac{\partial\mathcal{H}}{\partial q}=V+q/C(x)=V_\mathcal{R}(x)$ is the same as the drop across the equivalent resistor $\mathcal{R}$. The time averaged power dissipated at the resistor equals the time averaged power supplied by the thermal bath. Thus, from the resistor's perspective, the movement of the graphene ripple produces a constant source of average thermal power. If $\mathcal{R}$ were constant, the first term in Eq.~\eqref{eq8} would be Nyquist's noise power formula $\frac{T}{\mathcal{R}C}$. The exact result is shown in Fig.~\ref{fig3}(b), along with Nyquist's. Note the large power enhancement over the Nyquist result is due to the rate of change in resistance.

We confirmed these predictions by numerical simulation of Eqs.~\eqref{eq1} using the following: $T=0.5, \eta = 1, d = 10, I_0 = 0.0002$, and $T_e = 0.1$ (see Fig.~\ref{fig3} panels for other parameters and alterations). To account for the graphene shape change we have $C_0$ fall from 5 to 1 as V  increases from 1 to 10. To ensure numerical convergence, simulations were averaged over 10 million time steps and 1 million realizations. Particle position $x$ and charge at capacitor $q$ fluctuate with time as depicted in Figs.~\ref{fig3}(c) and \ref{fig3}(d). We separately calculate the two average power terms for the half cycle $\dot{q}>0$, in which current flows counterclockwise through diode 2. Even in the half cycle, the two terms are equal. The average power (both generated and dissipated), along with Nyquist's prediction is shown in  Fig.~\ref{fig3}(e). The power is found to increase with bias voltage, similar to our experimental results. Using resistance and power data from the experimental measurements presented in Fig.~\ref{fig2}, we estimate a capacitance near 1 fF for the tip-graphene junction.

The exact thermal power formula differs from the celebrated Nyquist in another significant way. The power includes contributions from the Brownian motion of the graphene ripple, not just that of the electrons. As a result, the double-well potential introduces a new time scale, which is the barrier crossing rate. This gives rise to very low frequency oscillations, which were characterized in our earlier study \cite{xu14}. To illustrate this, the average power spectral density for the power dissipated in the resistor is plotted using two different velocity relaxation times of 1 and 10, as shown in Fig.~\ref{fig3}(f). The total power dissipated is the same; a reduction in the barrier crossing rate redistributes power to lower frequencies, thereby adding technological value, as previously discussed by L\'opez-Su\'arez et al \cite{lop11}. 

In summary, we have studied the thermal fluctuations in freestanding graphene membranes using point-mode scanning tunneling microscopy. After disabling the STM feedback circuit, a displacement current was measured. We modeled the ripple closest to the STM tip as a Brownian particle in a double well potential. When the graphene moves, charge must flow through the circuit and perform electrical work. Our model provides a rigorous demonstration that continuous thermal power can be supplied by a Brownian particle at a single temperature while in thermodynamic equilibrium, provided the {\em same amount} of power is continuously dissipated in a resistor. Here, coupling to the circuit allows electrical work to be carried out on the load resistor without violating the second law of thermodynamics. Nonequilibrium fluctuations due to extra noises \cite{doe94,gha17} or to different temperatures in electrical circuits \cite{gon19} will produce entropy and measurable deviations from detailed balance \cite{gha17,gon19}, and are worth investigating in freestanding graphene.

This project was supported by the Walton Family Charitable Support Foundation and NSF grant DMR-0215872. M.R.-G. was supported by the Burroughs Wellcome Fund and Simons Foundation via Simons Investigator Award 327939. This work has also been supported by the FEDER/Ministerio de Ciencia, Innovaci\'on y Universidades -- Agencia Estatal de Investigaci\'on grant MTM2017-84446-C2-2-R. L.L.B. thanks John Neu for fruitful discussions and Russel Caflisch for hospitality during a sabbatical stay at the Courant Institute.

\end{document}